\documentclass{aa}  
\usepackage{layouts}
\usepackage{booktabs}
\usepackage{graphicx}
\usepackage[final]{microtype}
\usepackage{upgreek}
\usepackage{pgf}
\usepackage{float}
\usepackage{textcomp}
\usepackage{siunitx}
\usepackage{listings}
\usepackage{pythonhighlight}
\usepackage{array}
\usepackage{amsmath} 
\usepackage{multirow}
\usepackage{txfonts}
\usepackage{tabularx}
\usepackage{amssymb} 
\usepackage{fancyvrb}
\usepackage[colorlinks=true,linkcolor=blue,citecolor=blue, urlcolor=blue]{hyperref}

\begin{document} 

\title{Nine lensed quasars and quasar pairs discovered through spatially-extended variability in Pan-STARRS}
\titlerunning{Extended variability search in Pan-STARRS}
\authorrunning{Dux et al.}

\subtitle{}

\author{Frédéric~Dux\inst{1}, 
Cameron~Lemon\inst{1}, 
Frédéric~Courbin\inst{1}, 
Favio~Neira\inst{1, 2}, 
Timo~Anguita\inst{2, 3}, 
Aymeric~Galan\inst{1,4},
Sam~Kim \inst{6},
Maren~Hempel \inst{2,5},
Angela~Hempel \inst{2,5},
Régis~Lachaume \inst{5,6}
}

\institute{\email{frederic.dux@epfl.ch},\\
Institute of Physics, Laboratoire d’Astrophysique,  
École Polytechnique Fédérale de Lausanne (EPFL), 
Observatoire de Sauverny, CH-1290 Versoix, Switzerland
\and
Instituto de Astrofisica, Facultad de Ciencias Exactas, Universidad Andres Bello, Av. Fernandez Concha 700, Las Condes, Santiago, Chile 
\and
Millennium Institute of Astrophysics, Monseñor Nuncio Sotero Sanz 100, Oficina 104, 7500011 Providencia, Santiago, Chile 
\and
Technical University of Munich, TUM School of Natural Sciences, Department of Physics, James-Franck-Stra\ss{}e~1, 85748 Garching, Germany 
\and
Max Planck Institute for Astronomy, Königstuhl 17, 69117 Heidelberg, Germany 
\and
Pontificia Universidad Católica de Chile,
Avda. Libertador Bernardo O’Higgins 340, Santiago, Chile 
}

\date{Received July 29, 2023 at 23:59:59}
\abstract
{
We present the proof-of-concept of a method to find strongly lensed quasars using their spatially-extended photometric variability through difference imaging in cadenced imaging survey data. We apply the method to Pan-STARRS, starting with an initial selection of 14 107 \textit{Gaia} multiplets with quasar-like infrared colours from \textit{WISE}. We identify 229 candidates showing notable spatially-extended variability during the Pan-STARRS survey period. These include 20 known lenses, alongside an additional 12 promising candidates for which we obtain long-slit spectroscopy follow-up.
This process results in the confirmation of four doubly lensed quasars, four unclassified quasar pairs and one projected quasar pair. Only three are pairs of stars or quasar+star projections, the false positive rate is thereby 25\%. The lenses have separations between 0.81\arcsec and 1.24\arcsec and source redshifts between $z=1.47$ and $z=2.46$. Three of the unclassified quasar pairs are promising dual quasars candidates with separations ranging from 6.6 to 9.3 kpc.
We expect that this technique will be a particularly efficient way to select lensed variables in the upcoming Rubin-LSST, which will be crucial given the expected limitations for spectroscopic follow-up.
}
\maketitle

\section{Introduction}

Gravitationally lensed quasars span a vast range of astrophysical and cosmological applications. Among them, time-delay cosmography has the potential to address reliably the so-called Hubble tension \cite[e.g.][]{Verde2019}. 
The method relies on the photometric variability of quasars, which enables the measurement of the time delays between the lensed images, a quantity that is directly - and inversely - proportional to the Hubble constant \citep{Refsdal1964}. 
Currently, only seven lensed quasars have been used \citep{wong2020,shajib2020}, in part because of the paucity of known lensed quasars with sufficient variability. This scarcity has triggered a variety of searches, in recent wide-field datasets, for lensed quasars and the number of systems has doubled in the last few years alone \citep[e.g.][]{chan2022, lemon2023}. These searches have also generated useful by-products, such as dual quasars, i.e. physically associated quasars that are thought to be the progenitors of supermassive binary black holes responsible for the nano-hertz gravitational wave background \citep[e.g.][]{Mingarelli2019}, and a crucial element for understanding galaxy evolution \citep[e.g.][]{Kormendy2009}. 
On the other hand, as with any search for rare cosmic objects, contaminants often outnumber the targeted objects by several orders of magnitude. Among the most common contaminants are quasar-star projections and star-forming galaxies~\citep[e.g.][]{treu2018}.
Such contaminants typically feature only one or no variable point sources, so looking for variability in more than one point source -- hereafter \textit{extended variability} -- can be highly effective in removing contaminants, as originally suggested by ~\citet{Kochanek2006}. 

\citet{Lacki2009} perform such a search with difference imaging in the SDSS Supernovae Survey region, recovering known lensed quasars with a false positive rate of $\sim$ 1 in 4000. 
\citet{Kostrzewa2018} present the first variability-selected lensed quasar discovery. 
This was not done through difference imaging (as in the present work), but through similar variability features in their light curves, using the 3-day-cadence 700 square degrees field of the Optical Gravitational Lensing Experiment~\citep[OGLE]{udalski1992}. 
Their initial selection required quasar-like mid-infrared colours, and two candidates were confirmed as quasar-star projections. 
\cite{Chao2020} apply a difference-imaging-based search to the Hyper Suprime Cam (HSC) Transient Survey data, designed to effectively identify quadruply-lensed quasars. 
However, application to pre-selected variables only reduces the candidate list by $\sim$20 per cent, highlighting the need for additional visual inspection and further analysis \citep{Chao2021}.
\citet{Lemon2020} extract the Dark Energy Survey multi-epoch photometry of lensed quasars, quasar pairs, and quasar-star projections, and show that a simple cut in the less variable point source component of each system is indeed an effective way to recover all known lenses in the considered footprint while removing 94 percent of contaminants. 

The Pan-STARRS $3\pi$ survey~\citep{panstarrs, panstarrsimageprocessing} offers an unprecedented dataset for lensed quasar searches, having imaged 3/4 of the sky on average 7.6 times~\citep{Magnier_2013} in each of 5 filters between 2009 and 2012, with further observing epochs between 2012 and 2014. Furthermore, \citet{oguri2010} (hereafter OM10) predicts just under 2000 lensed quasars should be found in the Pan-STARRS dataset, to be compared with only $\sim$300 known systems as of today.\footnote{From a compilation of the literature, see also \citet{lemon2019a}.}

In this paper, we present a difference imaging algorithm and its application to lensed quasar candidate identification in Pan-STARRS. We demonstrate the algorithm's efficacy in identifying lensed quasar candidates, discuss the by-products of our search, and explore potential applications for future surveys. 
We first describe the details of the difference imaging method used herein in Section~\ref{sec:diffimg}.
Next, we describe the available datasets and selection of candidates in Section~\ref{sec:candidateselection}. 
In Section~\ref{sec:followupresults}, we present follow-up imaging and spectroscopy and a discussion on individual systems. 
We draw conclusions in Section~\ref{sec:conclusions}.

In all this work we assume a flat $\Lambda$CMD cosmology with $H_0=70{\rm \,km \,s^{-1}\, Mpc^{-1}}$ and $\Omega_m=0.3$.

\begin{figure*}[ht!]
\centering
\input{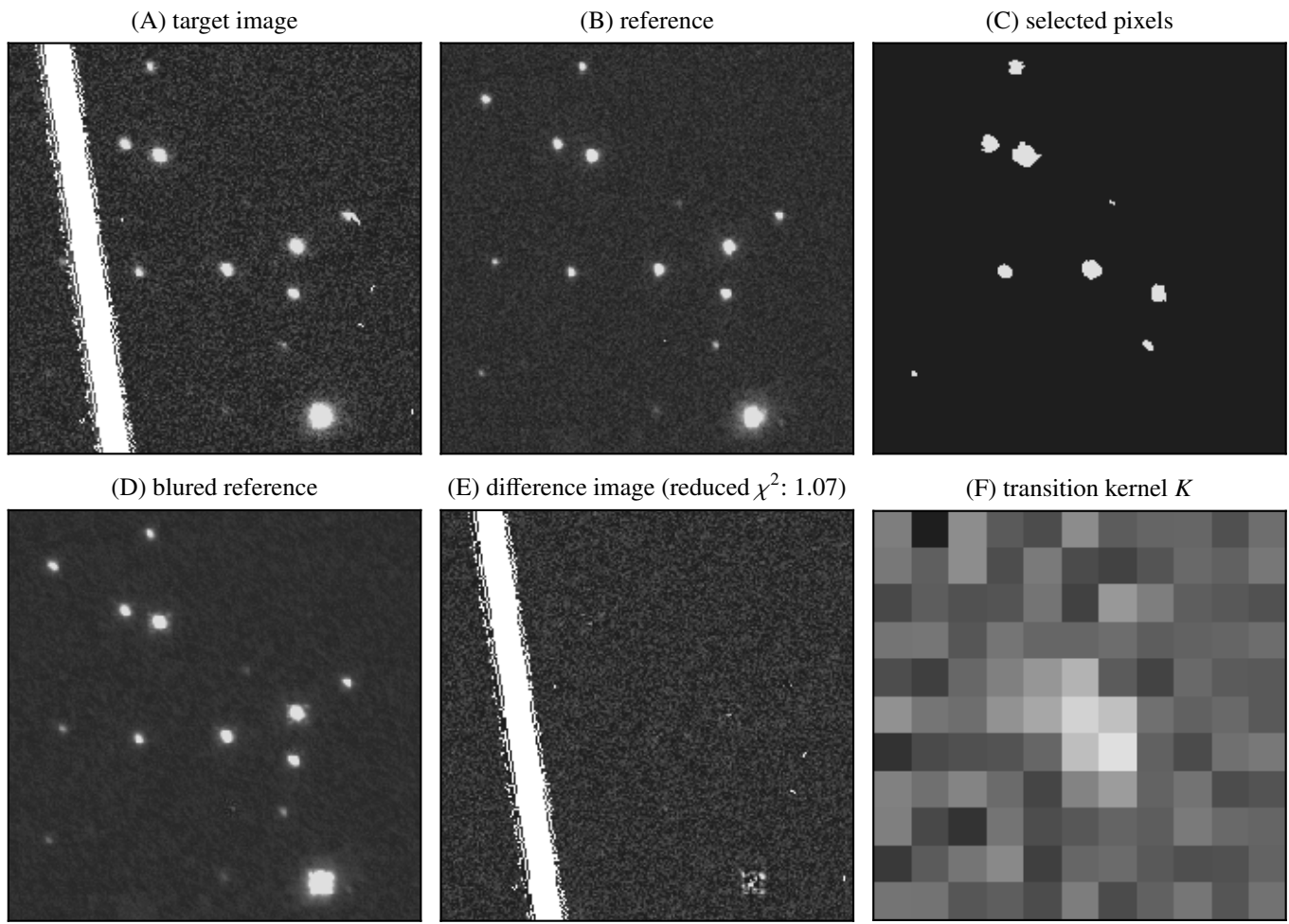}
\vspace{-0.5cm}
\caption{Illustration of the difference-imaging process using a 65\arcsec\ stamp of Pan-STARRS data. 
Panel (A) shows an image obtained at any given epoch ($I(\mathbf{x})$ in the text or {\it target image}) while panel (B) shows the best-seeing image of the time series, i.e. the reference image. 
Panel (C) shows the pixels that are used in the least square procedure. Note that saturated pixels are not considered. The blurred reference image is show in panel (D) 
and panel (E) shows the residuals after subtracting the blurred reference from the target image. 
The kernel used to blur the reference image is shown in panel (F). In the present case it is simply a low-pass filter coupled to a small translation accounting for the shift between the target and reference images.
}
\label{fig:example_difference}
\end{figure*}

\section{Difference imaging pipeline}
\label{sec:diffimg}

Detecting photometric variations of astronomical objects implies obtaining images at different times and therefore of different quality. The seeing conditions and sky level in particular are bound to vary from 
one image to the next. In order to detect an intrinsic difference in the total 
luminosity of a given object, the simplest strategy is to subtract a reference 
image from all other images in the set. In an ideal situation, the only signal rising above 
the noise in such a difference-image would be that of the sought 
intrinsic variability. This will naturally never be the case because of the varying
sky seeing and sky brightness. Hence a proper procedure needs to be devised to account for the different point spread functions (PSF) and overall brightnesses of the images. 

In this section, we first introduce the main idea of difference imaging and introduce {\it variability maps}, which are the stacks of difference images used to assess extended variability. Lastly, we benchmark the method on real data. We make the resulting pipeline available online as a Python package.\footnote{\url{https://github.com/duxfrederic/diffpanstarrs}}.

\subsection{Basics of difference imaging}
\label{sec:method-main}

Given a 2-dimensional image $I(\mathbf{x})$, and a reference image $R(\mathbf{x})$ of the same size, we adopt the following simple model that blurs $R(\mathbf{x})$ so it matches the resolution of $I(\mathbf{x})$:
\begin{equation}
\label{eq:conv}
R_\mathrm{blurred}(\mathbf{x}) = \int \mathrm{d}\mathbf{x'} K(\mathbf{x}, \mathbf{x'}) \, R(\mathbf{x'}) + B,
\end{equation}
where the reference image $R$ is convolved with the kernel $K$ that minimizes the following~$\chi^2$:
\begin{equation}
\label{eq:chi2}
\chi^2 = \sum_{i=1}^{n}\frac{\big[  I(x_i)- R_\mathrm{blurred}(x_i) \big]^2}{\sigma_R^2(x_i)+\sigma_I^2(x_i)},
\end{equation}
with $n$ the total number of pixels in the image. Usually the reference image is the one with the best seeing such that $K$ can be a low-pass filter as detailed in Appendix~\ref{sec:refimage}.
The constant term $B$ in Eq.~(\ref{eq:conv}) accounts for the background level of the sky, and is assumed to be constant across the field of view. 
Additional terms (e.g. polynomials in the image coordinates) can potentially  be added to 
better fit any significant light gradient, however the considered fields are small enough to make this unnecessary. 
Next, the kernel is decomposed onto a set of basis functions whose weights can be tuned to minimize Eq.~(\ref{eq:chi2}).
Choosing Gaussian basis functions yields the approach taken by \citet{Alard1998} for the analysis of Galactic microlensing light curves. 
We adopt instead the more robust (albeit slower) approach of \citet{Bramich2008} and we
set the kernel to be a sum of delta distributions, namely a pixelated kernel. With such a kernel, the convolution~(\ref{eq:conv}) reads

\begin{equation}
\label{eq:convpixel}
R_\mathrm{blurred}(x_i) = \sum_{j=1}^{N^2} K_j R_{ij} + B.
\end{equation}
With $R_{ij}$ the $j^{\rm \, th}$ pixel of the reference image in the neighborhood centered on the $i^{\rm \, th}$ pixel, $K_j$ are the values of the pixelated kernel, and $N$ is the extent of the kernel.
Now we can cast this last equation as a least squares problem by requesting that the blurred reference should be equal to the target image.
\begin{equation}
\label{eq:bigmatrix}
\renewcommand{\arraystretch}{1.5}
R_\mathrm{blurred} =
\begin{bmatrix} 
R_{1\,1} & \dots  & R_{1\,N^2} &1\\
\vdots & \ddots & & \vdots \\
R_{n\,1} &        & R_{n\, N^2}  & 1
\end{bmatrix}
\renewcommand{\arraystretch}{1.}
\begin{bmatrix}
K_1 \\
\vdots \\
K_{N^2} \\
B
\end{bmatrix} 
\overset{!}{=} I.
\end{equation}
Each column of the matrix corresponds to a translation of the reference image, 
and each line corresponds to a pixel of the reference image.
Eliminating bad pixels from the problem is thereby very easy, 
we just need to remove the corresponding line from the matrix equation.
We also further refine the problem by dividing both the reference image and the target image with their respective noisemaps before building the matrix. Finally, we can use the closed-form solution of Eq.~(\ref{eq:bigmatrix}) to obtain both the kernel values $K_i$ and background $B$ which minimize the $\chi^2$ of Eq.~(\ref{eq:chi2}). This is much faster than using an iterative minimization algorithm.

\subsection{Variability maps}
\label{sec:method-varmap}

Provided the difference images at all observing epochs have been perfectly computed, they can be combined  into a {\it variability map} that we define as follows:
\begin{equation}
I_\mathrm{var} = \sum_i\frac{\big\vert  R_{\mathrm{blurred}\to i} - I_i\big\vert }{\sigma_{R,\,\mathrm{blurred}\to i    }^2+\sigma_{I_i}^2}
\end{equation}
where $R_{\mathrm{blurred}\to i}$ is the blurred version of the reference image such that it matches the PSF and background of the $i^{\mathrm{th}}$ target image.
In words, the variability map is the sum of the absolute difference images weighted by the corresponding noise maps. Note that we also adapt the resolution of the noisemap of the reference image, see Appendix~\ref{sec:noisemap_diffimg}.

The overall difference-imaging process is illustrated using Pan-STARRS data in Fig.~\ref{fig:example_difference}.

\subsection{Tests with Pan-STARRS data}

We perform two types of tests of the method on real data. First, we apply it to known lensed quasars. Starting from the 191 lensed quasars in~\cite{lemondatabase} that fall in the Pan-STARRS footprint, we find that 133 show extended variability. This is encouraging performance, even though
34 of the variability maps were not reliable due to the high brightness of the targets themselves. In these cases, the transition kernel $K$ is not properly constrained from the fainter stars in the field, causing a poor subtraction of the PSFs of the candidate and leaving non-PSF-like residuals.
This experiment allows us to set the threshold of reliability of our method
at a \textit{Gaia} G-mag of 17, given the magnitude distribution of PSF stars available in practice.

Second, we examine the case of apparent quasar+star pairs, which are the most common contaminants to lens searches. Fortunately, the fraction of variable quasars is much higher than the fraction of variable stars, as the majority of quasars tend to vary by more than 0.1 mag over two years~\citep{Cimatti1993}.
To benchmark the correct rejection of stars, we build a small selection of quasar+star systems by cross-matching the MILLIQUAS \citep{flesch2021million} and Gaia EDR3 \citep{gaiaedr3} catalogues and imposing additional restrictions: we request a large enough proper motion significance (PMSIG, see \citealt{Lemon2018} for the definition) of the companion to make sure we are not including another quasar, and a small \textit{Gaia} $G$-magnitude difference between the two.
Ideally, the resulting systems should all produce variability with at most one PSF visible. This is achieved for 164 of the 170 test-systems. 
Among the remaining 6, 3 do not have sufficient Pan-STARRS coverage to perform difference imaging, 
while for the other 3 the difference imaging performs poorly as there are too few reference stars available even when using large cutouts. In such cases the variability maps wrongly show variability.  
This goes to show that the method is able to efficiently reject quasar+star pairs, although care must still be taken to assess the quality of the difference imaging. 

\begin{figure*}[!ht]
\centering
\resizebox{\linewidth}{!}{
 \input{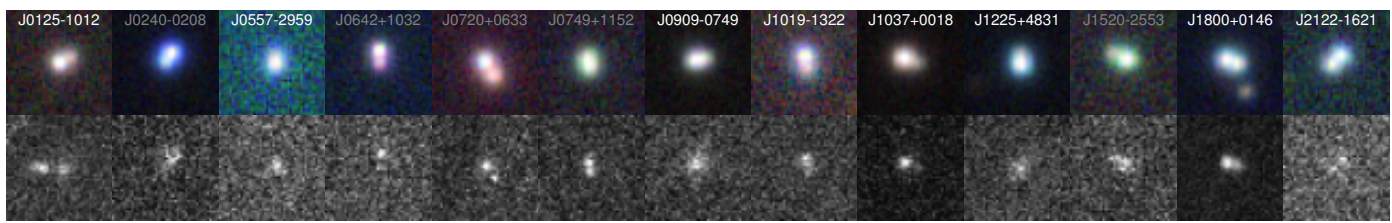}
}
\caption{Pan-STARRS \textit{gri} views (top) and \textit{r+i} variability maps (bottom) of the candidates selected for spectroscopic follow up.
         Extended variability is, albeit sometimes barely raising above the noise, still well visible for all the candidates.
         Those that are found to be false positives are displayed in gray, whereas the quasar pairs are displayed in white. 
         Note that the variability map of PS$\,$J0125$-$1012 has an artifact.}
\label{fig:news_mosaic}
\end{figure*}

\section{Candidate Selection}
\label{sec:candidateselection}

To maximize the chances of lens discovery we first build a catalogue of sources that are likely to contain a quasar. To do this, we combine the
\textit{Gaia} \citep{gaiadr2} and \textit{WISE} \citep{Wright2010} catalogues.
We start with all the \textit{WISE} detections that are red enough, 
that is the \textit{WISE} color fulfils:
\[
W_\mathrm{3.4 \upmu m} - W_\mathrm{4.6 \upmu m} > 0.7.
\]
This criterion is stricter than that used by~\citet{lemondatabase}, but helps reducing the number of candidates to a manageable number for this proof of concept at the cost of losing bluer candidates.
Then, among the red \textit{WISE} detections  we select only those that have a first \textit{Gaia} detection within 3\arcsec, and a second \textit{Gaia} detection between 0.8\arcsec\ and 3\arcsec\ away from the first one. All the detections with a PMSIG larger than 3 are discarded. Also eliminated are all the detections within 15$^\circ$ of the galactic plane to reduce the likelihood of selecting stars.
Given the above criteria, our final selection consists of $14\,591$ objects. 

Next, we make use of all the available Pan-STARRS cutouts in the $r$ and $i$ band of the objects selected above.
The choice of the $r$ band is motivated by the fact that quasars are more variable in bluer bands~\citep[e.g.][]{macleod2010}, but the $g$ band typically comes with harder-to-model PSFs.
The $i$ band is also included to patch the regions where the $r$ band is not available. 
The analysis could not be performed for 484 candidates due to insufficient data, as fewer than two epochs were available for these.
Consequently, a total of $14\,107$ variability maps are generated using our difference-imaging method. Out of these, $9\,692$ maps are discarded due to the absence of any discernible signal above the background noise level. This leaves $4\,415$ maps requiring visual inspection. 
We exclude $3\,300$ more that show a single PSF-like structure in the variability maps but no extended variability.
For the remaining maps displaying extended variation, we make sure that the stars in the rest of the field are indeed properly subtracted. This is not the case for 720 more maps, in which the difference-imaging procedure fails. 
This is usually caused by a lack of bright enough stars in field, leading to a poor constraint of the transition kernel~$K$, or other defects: see Appendix~\ref{sec:broken}.
Lastly, we inspect the Pan-STARRS \textit{gri} color stamps of each candidate, giving priority to those showing consistent color across PSFs. We exclude 166 more objects which evidently do not look like lenses.
Ultimately we identify a total of 229 variable objects, including 20 known lenses. From this subset, we further narrow down our selection to 14 particularly promising candidates for spectroscopic follow-up.

\begin{figure*}
  \input{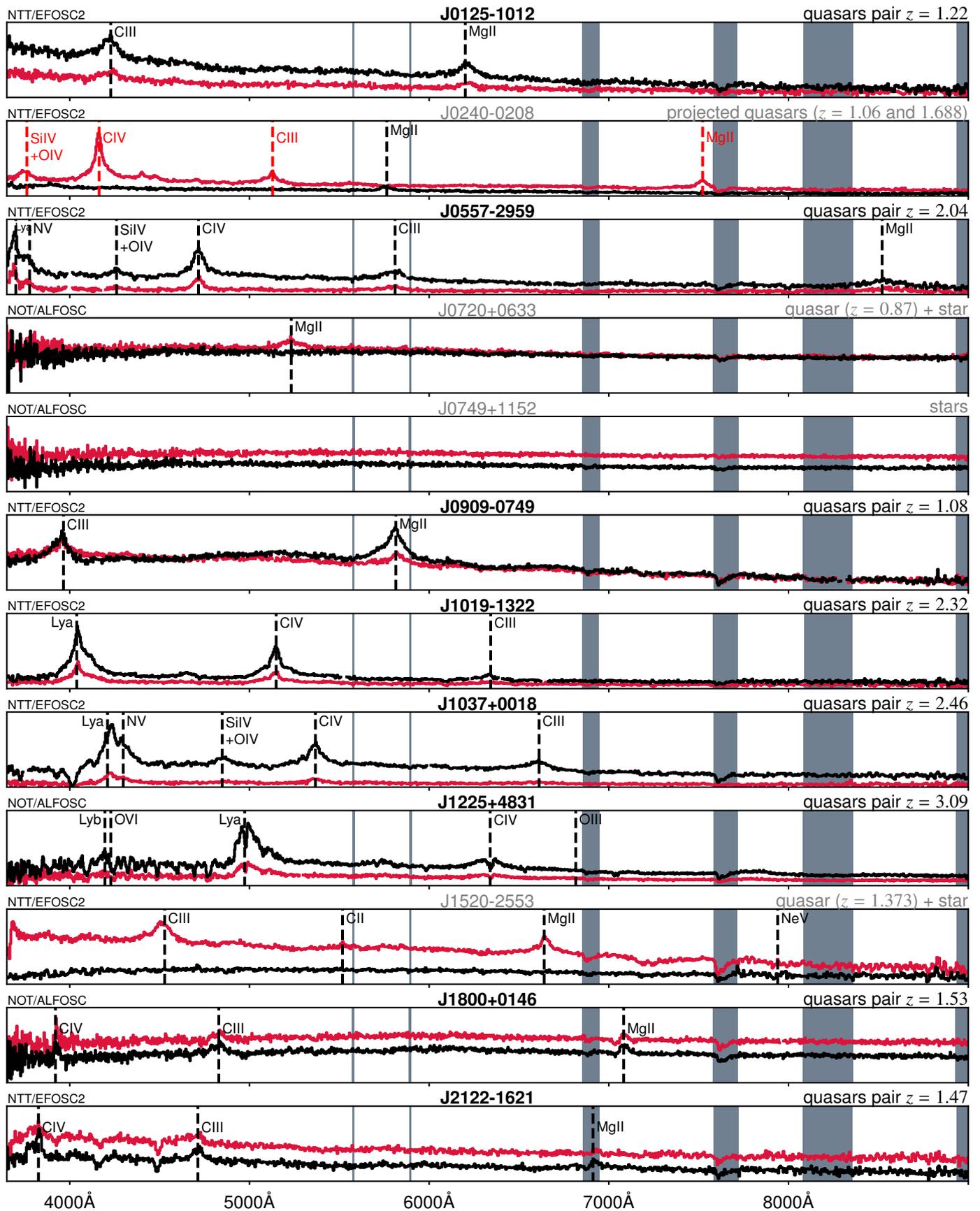}\vspace{-0.25cm}
  \caption{Long slit spectroscopy of double-type candidates performed using EFOSC2 on the ESO/NTT or ALFOSC on the Nordic Optical Telescope. The candidates are arranged in ascending order of right ascension, with lenses or quasar pairs written in bold and stars or projected quasars in gray. Shaded regions indicate wavelength ranges affected by atmospheric absorption. The spectra were reliably de-blended using the procedure outlined in Section~\ref{sec:followupspectro}.} 
  \label{fig:all_spectras}
\end{figure*}

\section{Follow-up observations and results}
\label{sec:followupresults}

\subsection{Spectroscopy}
\label{sec:followupspectro}

Eight systems were observed with grism \#13 with the ESO Faint Object Spectrograph and Camera 2 (EFOSC2) on the ESO New Technology Telescope (NTT), over two
runs in October 2020 and January 2021\footnote{Timo Anguita, 106.218K.001 and 106.218K.002.}, providing a dispersion of 2.77\si{\angstrom} per pixel. Each object received between 900 and $1\,200$ seconds of exposure time. Four other systems were observed with grism \#4 and the Alhambra Faint Object Spectrograph and Camera (ALFOSC) on the Nordic Optical Telescope (NOT) in April 2021\footnote{Cameron Lemon, 63-106.}, providing a dispersion of 3.3\si{\angstrom} per pixel. Each object received between 900 and $1\,200$ seconds of exposure time.

The general spectral reduction is carried out exactly as described in \cite{lemon2023}. 
The key steps are as follows. Each 2D long-slit spectrum image is bias-subtracted, and cosmic rays are masked using a Laplacian threshold method. 
The threshold is adjusted based on the brightness of the objects and a visual inspection of the cosmic ray mask. 
Next, the sky background is subtracted at each wavelength by determining the median value within the pixels surrounding the object trace, providing a sky-subtracted, debiased, cosmic-ray-masked 2D long-slit spectrum image. 
A corresponding 2D noise map is then generated, combining a Poisson noise map derived from the detector gain with the sky noise. 
Finally, a wavelength solution is established by fitting lines from a pre-observation arc exposure (HeNe or CuAr), and linearly shifted for each exposure to fit several major sky background emission lines. 
At each wavelength, a model of the 1D spatial profile of the two components is generated as two Moffat profiles separated by the Gaia separation, and integrated perpendicularly in the slit. 
The Moffat parameters as a function of wavelength are then determined by fitting this model to several bins across wavelength. 
Finally the flux of each component can be derived at each wavelength by using the best-fit Moffat parameters at each wavelength.

\begin{figure*}[t!]
    \centering
    \resizebox{1.\linewidth}{!}{
       \input{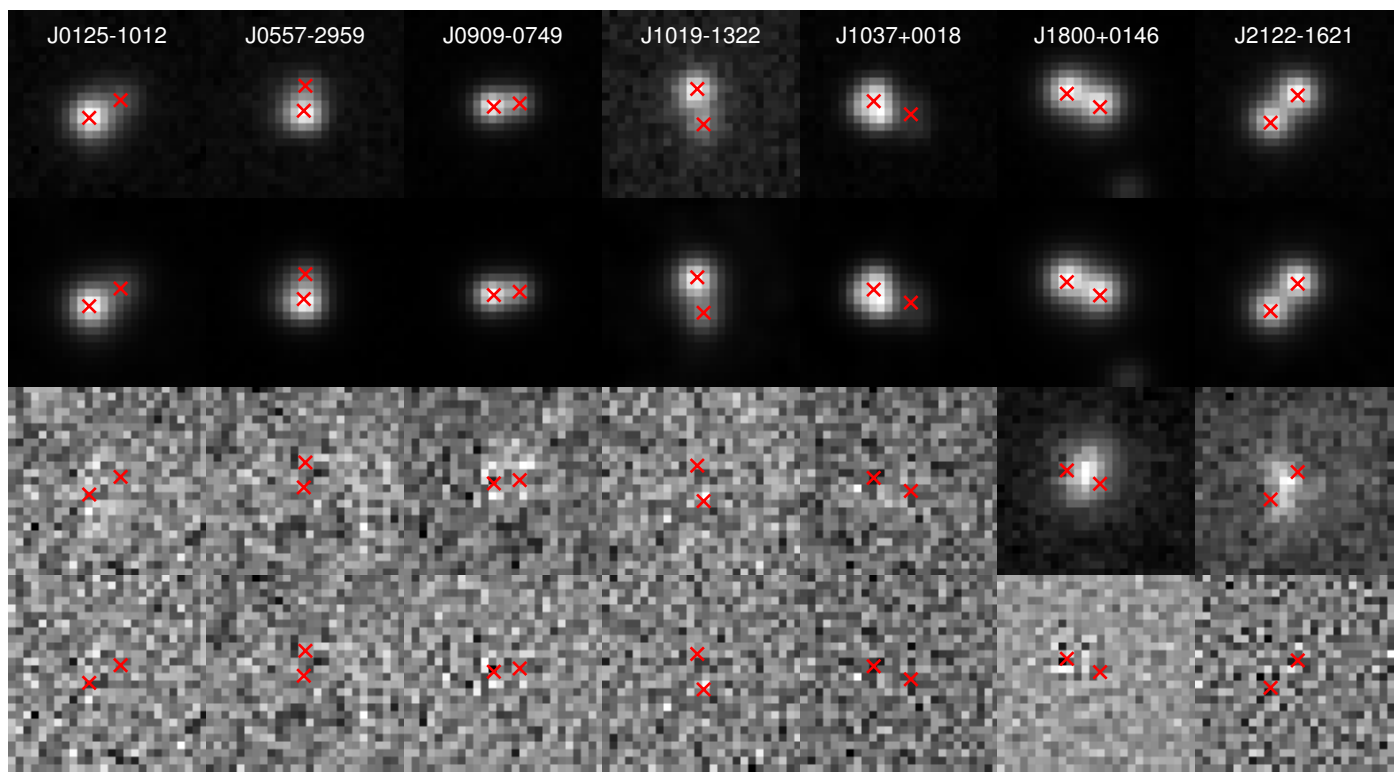}
    }
    \caption{\textit{First row:} MPIA~2.2-meter telescope $R_c$-band data (4$\times$320 seconds). \textit{Second row:} modeling of the systems using the STARRED algorithm. \textit{Third row:} residuals after subtractions of the point-source component of the STARRED model. \textit{Last row:} residuals after subtraction of the complete STARRED model (PSFs + background). 
    Each row shows a stack, but the data was jointly deconvolved. The red crosses indicate the positions of the PSFs, the separations of which are close to the Gaia separations down to 0.01\arcsec\ (Table~\ref{tab:news}). At this depth and filter, the lensing galaxy is detectable only in the two brightest objects. 
    \label{fig:2m2deconv}}
\end{figure*}

\subsection{Imaging}
\label{sec:imagingsetup}
The Pan-STARRS data are limited in depth when trying to detect any potential lens galaxy between the quasar pairs. Given the narrow separations and faint magnitudes of our candidates, we obtained deeper images of the 7 Southern double-type candidates in the $R_c$ band using the Wide Field Imager (WFI) on the Max-Planck-Institut für Astrophysik (MPIA)~2.2-meter telescope (2p2), in nights of excellent seeing between March and June 2022. We obtained a total exposure time of $4\times320$ seconds per target.

Due to the complex PSF in these new images, we build a PSF model oversampled by a factor of 3, using up to 6 stars in each field. Next, for each object we perform a joint deconvolution, fitting the positions and amplitudes of two point sources as well as a Starlet-regularized pixelated background. Because galaxies are known to be sparse in Starlet space, we expect the lensing galaxy to naturally emerge in the background, if detectable in the dataset. We use STARRED \citep{starred} to perform both the PSF modelling and the image deconvolution. The result is shown in Fig.~\ref{fig:2m2deconv}.

When deeper Legacy Survey data or Hyper Suprime-Cam Subaru Strategic Program (HSC, \citeauthor{Aihara:2021jwb}~\citeyear{Aihara:2021jwb}) data are available, we perform a similar analysis but we replace the pixelated background with an analytical Sersic~\citep{sersic} profile. The Legacy Survey and Pan-STARRS data are then jointly fitted, but the HSC images are analyzed on their own because of their vastly superior signal-to-noise.
We make the code developed for this analysis publicly available as well\footnote{\url{https://github.com/duxfrederic/lensedquasarsurveyor}}.

\begin{table*}
\centering
\renewcommand{\arraystretch}{1.3} 
\caption{Confirmed quasars pairs. Are given the separation between the two images from \textit{Gaia}, the spectroscopic redshift and the \textit{Gaia} G-magnitudes. 
When applicable, we also indicate the separation recovered from the models in Figs.~\ref{fig:2m2deconv} and \ref{fig:survey_fitted_images}.
}
\label{tab:news}
\begin{tabularx}{\linewidth}{lrrrrrrrr}
\toprule
Object & R.A. & Dec. & \textit{Gaia} (PSF) sep. & \textit{Gaia} G-mag & $z$  & Type & Referenced by \\
\midrule
PS~J0125$-$1012 & 21.3174 & -10.2082  & 1.12\arcsec (1.13\arcsec)  & 19.25, 20.61 & 1.22 & UQP & \citet{lemon2023}  \\
PS~J0240$-$0208 & 40.0768 & -2.1474 & 0.94\arcsec    & 18.70, 19.58 & 1.688, 1.08 & Projected pair   \\
PS~J0557$-$2959 & 89.3979 & -29.9933 & 0.81\arcsec (0.80\arcsec)  & 19.50, 20.56 & 2.04 & Lens &  \\
PS~J0909$-$0749 & 137.4946 & -7.8179 & 0.81\arcsec (0.82\arcsec) &  18.25, 19.05 & 1.08 & UQP &  \\
PS~J1019$-$1322 & 154.8066 & -13.3692 & 1.13\arcsec (1.13\arcsec) & 19.57, 20.41 & 2.32 & UQP &  \\
PS~J1037+0018 & 159.3665 & 0.3057 & 1.24\arcsec (1.23\arcsec) & 18.05, 19.99 & 2.46 & Lens &  \\
PS~J1225+4831 & 186.3278 & 48.5212 & 0.90\arcsec  &    18.00, 19.26 & 3.09 & UQP  & \citet{Minghao2023}  \\
PS~J1800+0146 & 270.0320 & 1.7767 & 1.12\arcsec (1.12\arcsec) & 18.39, 18.72 & 1.53 & Lens &  \\
PS~J2122$-$1621 & 320.6075 & -16.357 & 1.20\arcsec (1.21\arcsec) & 19.14, 19.26 & 1.47 & Lens &  \\
\bottomrule
\end{tabularx}
\end{table*}

\subsection{Discussion: quasar pairs}

\begin{figure}[b!]
    \centering
    \includegraphics[width=\columnwidth]{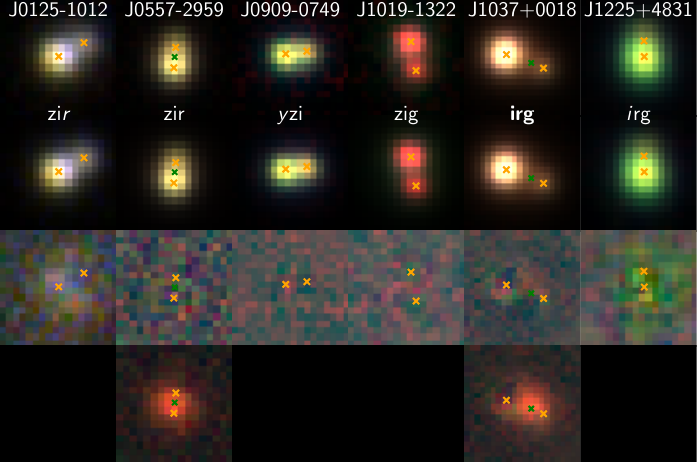}
    \caption{
    \textit{First row:} Legacy Survey, Pan-STARRS (italicized) or HSC (bolded) data. 
    \textit{Second row:} Models with 2 PSFs, and also an additional Sersic profile for J0557-2959 and J1037+0018.
    \textit{Third row:} Residuals after subtraction of the models.
    \textit{Fourth row:} When the model has a Sersic profile, residuals after subtraction of the PSFs only.
    The bands are arranged by wavelength (redder first), highlighting the redder hue of the lensing galaxies seen in the fourth row compared to the quasar pairs.}
    \label{fig:survey_fitted_images}
\end{figure}

Figure~\ref{fig:news_mosaic} shows the Pan-STARRS \textit{gri} 
stacks and variability maps of all our double-type candidates for which we have follow-up imaging, ordered by right ascension.
The variability of PS$\,$J0125$-$1012 clearly suffers from of an artifact in one of the cutouts, but we still select it given the general look of the Pan-STARRS \textit{gri} cutout. 
We show the acquired spectra in Fig.~\ref{fig:all_spectras}. 
Among the 12 double-type candidates, 8 are spectroscopic quasar pairs (similar spectra in both components), 1 is a projected quasar pair, 2 are quasar+star pairs, and 1 is a pair of stars. Below we tentatively classify the spectroscopic quasar pairs with comments referring to the spectra of Fig.~\ref{fig:all_spectras}, the 2p2 images of Fig.~\ref{fig:2m2deconv}, and the available survey data of Fig.~\ref{fig:survey_fitted_images}.

\paragraph{PS~J0125$-$1012 $\;-$}
The NTT/EFOSC2 resolved spectrum shows two quasars at redshift $z=1.22$ with very similar continua and broad emission lines.
We jointly fit the \textit{iz} Legacy Survey and \textit{rzy} Pan-STARRS bands. The data is well modelled with only two point sources, and the resulting image separation is 1.12\arcsec, compatible with the Gaia measurement. Although the reduced $\chi^2$ of the fit can be slightly improved by allowing for an additional Sersic profile, the improvement is not significant and the point sources are pushed further apart, breaking the compatibility with the Gaia separation.
Considering the similar 50 OM10 mock doubles, i.e. with source redshifts within $\Delta z = 0.1$ and image separations within 0.1\arcsec, their faintest lens galaxy magnitudes are $i\sim21.1$.
This should be detectable at the depth of our Legacy Survey + Pan-STARRS cutouts.
If this system turns out to be a dual quasar, the physical separation would be 9.3~kpc. 
We suggest this system is a dual quasar, but designate this system as an Unclassified Quasar Pair (UQP) until deeper high-resolution imaging can be obtained.

\paragraph{PS~J0557$-$2959 $\;-$}
The NTT/EFOSC2 resolved spectrum shows two quasars at redshift $z=2.04$ with identical continua and broad emission lines. Even though our shallower $r$-band 2p2 data is well fitted with two PSFs, the same fit of Legacy Survey data shows a clear sign of a lensing galaxy in the $z$-band data. This residual subsequently disappears when including a Sersic profile, which falls between the quasars. We also note that the PSF separation is 0.783\arcsec when fitting without a galaxy, and 0.808\arcsec when fitting with a galaxy, the latter of which agrees better with the Gaia separation of 0.810\arcsec, providing more evidence for the existence of the lensing galaxy. We therefore conclude that this system is a lensed quasar.

\paragraph{PS~J0909$-$0749 $\;-$}
The NTT/EFOSC2 resolved spectrum shows two quasars at redshift $z=1.08$ with distinct differences in the ratio of the continuum and broad lines. There is no evidence for a lensing galaxy in the WFI 2p2 imaging. The simultaneous fit of Legacy Survey \textit{iz} and Pan-STARRS \textit{grizy} data with two point sources is good, and does not significantly improve when adding a Sersic profile. The Gaia separation is also recovered without the need for a model containing a galaxy. Moreover, the 38 similar OM10 mocks can have galaxies as faint as $i\sim21.2$, which should be detectable.
Given the slight differences in spectra and lack of lensing galaxy, we suggest this system is a dual quasar with physical separation 6.6~kpc.
However we conservatively designate this system as an UQP, needing deeper high-resolution imaging.

\paragraph{PS~J1019$-$1322 $\;-$}
The NTT/EFOSC2 resolved spectrum shows two quasars at redshift $z=2.32$ with very similar emission line profile shapes. The flux ratio across wavelength is $\sim$2.2, with some small variations around the emission lines. There is no evidence for a lensing galaxy in the WFI 2p2 imaging. 
Similarly, the simultaneous fit of the Pan-STARRS \textit{grizy} and Legacy Survey \textit{griz} data with two point sources does not leave significant residuals, and the Gaia separation is recovered as well.
Similar mocks from the OM10 catalogue can have galaxy magnitudes as faint as $i\sim22.6$. We suggest this system is a lensed quasar given the similarity of the spectra, however a deep image detecting the lensing galaxy is still needed. We designate this system as an UQP.

\paragraph{PS~J1037+0018 $\;-$}
The NTT/EFOSC2 resolved spectrum shows two very similar quasars at redshift $z=2.46$. The flux ratio is approximately 5.5 across all wavelengths, and both spectra show a damped Lyman-alpha system. The available HSC images are poorly fit with two point sources, but the subsequent introduction of a Sersic profile produces much cleaner residuals. Subtracting the two point sources then reveals the lensing galaxy close to the fainter point source.
We designate this system as a lens.

\paragraph{PS~J1225+4831 $\;-$}
This system has also been discovered by \citet{Minghao2023}, who obtained Large Binocular Telescope
LUCI1 imaging with an effective PSF of 0.35\arcsec\ (FWHM). The images show no evidence for a lensing galaxy, and they further note a possible flux ratio difference between the H$\beta$ and \textsc{[Oiii]} lines. 
We present resolved NTT/EFOSC2 spectra of the system, which show two quasars at redshift $z=3.09$ with a reasonably constant flux ratio of 3 across wavelength. Similar systems in the OM10 mock catalogue have lensing galaxies as faint as $i\sim23.7$. One possible explanation of these observations is that the lensing galaxy falls very close to one of the images and is being fit by the PSF subtraction of \citet{Minghao2023}, and the fainter component's spectrum is too noisy to conclude incompatibility with the brighter object's spectrum.
Additionally, jointly fitting the available Pan-STARRS and Legacy Survey data produces slightly prominent residuals with a 2-PSFs model, but the color of the said residuals matches that of the point sources. We thus attribute the poor fit to a defect in our PSF model, and thereby cannot find evidence for a lensing galaxy either. If this system were a dual quasar, its physical separation would be 6.87 kpc.
We designate this system as an UQP, needing deeper imaging to confirm the presence or absence of the lensing galaxy.

\paragraph{PS~J1800+0146 $\;-$}
The NOT/ALFOSC resolved spectrum shows two very similar quasars at redshift $z=1.53$. The flux ratio is roughly one across wavelength. Here the 116 similar OM10 mocks suggest a lensing galaxy as faint as $i\sim21.6$, which is relatively bright and was blindly picked up by starlets in the 2p2 data. The Gaia separation of 1.12\arcsec\ is also recovered by the fit. We designate this system as a lens.

\paragraph{PS~J2122$-$1621 $\;-$}
The NTT/EFOSC2 resolved spectrum shows two very similar quasars at redshift $z=1.47$, with a flux ratio of $\sim1.2$ across wavelength. Just like in the case of PS~J1800+0146, the lensing galaxy could be fitted blindly with STARRED in the 2p2 data. The Gaia separation of 1.20\arcsec was recovered as well. We designate this system as a lens.

\section{Conclusions}
\label{sec:conclusions}
Our search using difference imaging in Pan-STARRS has led to the identification of four new doubly lensed quasars and four Unclassified Quasar Pairs that are either lenses or dual quasars, and one projected quasar pair.
Notably, all of the lenses have narrow separation between 0.81\arcsec\ and 1.24\arcsec, and three have source redshifts above $z=1.5$. 
This distribution is thereby starting to fill the gap pointed out by \citet{lemon2023}, where many small separation high redshift doubles are missing.
Furthermore, \citet{Lemon2020} note that there is a quad bias when variability is used to select lensed quasars, which can be understood by the fact that the more variable quasars are the least massive ones~\citep{macleod2010}: these have fainter absolute magnitudes and would need to be highly magnified to reach a detectable threshold.
Thereby searching for lensed quasars through their variability introduces a high-magnification bias, and is complementary to traditional search methods for the construction of a magnitude-limited lensed quasar catalogue.
So, given the bias for strong magnification that our technique introduces, it may seem surprising that we do not find more quads. 
This is simply due to the fact that all quads were already found at the Gaia limit \citep{Lemon2018,Delchambre2019,lemon2023}.
As we extend the search to beyond the Gaia limit, we can expect that this time, quads will be detected at an increased rate.

In conclusion, we have presented a difference imaging technique and applied it to Pan-STARRS imaging with a Gaia+WISE pre-selection, focusing on detecting extended variability as an efficient method for the selection of lensed quasars. 
This work underlines the potential of extended variability and difference imaging as an effective method for lensed quasar selection, with great promise of uncovering highly magnified lenses in future surveys such as the forthcoming Vera Rubin Observatory's Legacy Survey of Space and Time ~\citep[LSST]{lsst}. 
Indeed, considering the anticipated scarcity of spectroscopic resources in the future relative to the vast number of candidates expected to be uncovered by LSST, attaining a high level of purity in candidate selection through their  variability will become increasingly important.

\section{Acknowledgments}

This program is supported by the Swiss National Science Foundation (SNSF) and by the European Research Council (ERC) under the European Union’s Horizon 2020 research and innovation program (COSMICLENS: grant agreement No 787886).
TA acknowledges support from the Millennium Science Initiative ICN12\_009 and the ANID BASAL project FB210003

\paragraph{2p2:} The 2p2 imaging was made possible by an agreement between MPIA and EPFL, the ESO program ID is 0108.A-9005(A).

\paragraph{Pan-STARRS:} The Pan-STARRS1 Surveys (PS1) and the PS1 public science archive have been made possible through contributions by the Institute for Astronomy, the University of Hawaii, the Pan-STARRS Project Office, the Max-Planck Society and its participating institutes, the Max Planck Institute for Astronomy, Heidelberg and the Max Planck Institute for Extraterrestrial Physics, Garching, The Johns Hopkins University, Durham University, the University of Edinburgh, the Queen's University Belfast, the Harvard-Smithsonian Center for Astrophysics, the Las Cumbres Observatory Global Telescope Network Incorporated, the National Central University of Taiwan, the Space Telescope Science Institute, the National Aeronautics and Space Administration under Grant No. NNX08AR22G issued through the Planetary Science Division of the NASA Science Mission Directorate, the National Science Foundation Grant No. AST-1238877, the University of Maryland, Eotvos Lorand University (ELTE), the Los Alamos National Laboratory, and the Gordon and Betty Moore Foundation.

\paragraph{HSC:}
The Hyper Suprime-Cam (HSC) collaboration includes the astronomical communities of Japan and Taiwan, and Princeton University. The HSC instrumentation and software were developed by the National Astronomical Observatory of Japan (NAOJ), the Kavli Institute for the Physics and Mathematics of the Universe (Kavli IPMU), the University of Tokyo, the High Energy Accelerator Research Organization (KEK), the Academia Sinica Institute for Astronomy and Astrophysics in Taiwan (ASIAA), and Princeton University. Funding was contributed by the FIRST program from Japanese Cabinet Office, the Ministry of Education, Culture, Sports, Science and Technology (MEXT), the Japan Society for the Promotion of Science (JSPS), Japan Science and Technology Agency (JST), the Toray Science Foundation, NAOJ, Kavli IPMU, KEK, ASIAA, and Princeton University.  

\paragraph{Legacy Survey:}
The Legacy Surveys consist of three individual and complementary projects: the Dark Energy Camera Legacy Survey (DECaLS; Proposal ID \#2014B-0404; PIs: David Schlegel and Arjun Dey), the Beijing-Arizona Sky Survey (BASS; NOAO Prop. ID \#2015A-0801; PIs: Zhou Xu and Xiaohui Fan), and the Mayall z-band Legacy Survey (MzLS; Prop. ID \#2016A-0453; PI: Arjun Dey). DECaLS, BASS and MzLS together include data obtained, respectively, at the Blanco telescope, Cerro Tololo Inter-American Observatory, NSF’s NOIRLab; the Bok telescope, Steward Observatory, University of Arizona; and the Mayall telescope, Kitt Peak National Observatory, NOIRLab. Pipeline processing and analyses of the data were supported by NOIRLab and the Lawrence Berkeley National Laboratory (LBNL). The Legacy Surveys project is honored to be permitted to conduct astronomical research on Iolkam Du’ag (Kitt Peak), a mountain with particular significance to the Tohono O’odham Nation.
NOIRLab is operated by the Association of Universities for Research in Astronomy (AURA) under a cooperative agreement with the National Science Foundation. LBNL is managed by the Regents of the University of California under contract to the U.S. Department of Energy.
This project used data obtained with the Dark Energy Camera (DECam), which was constructed by the Dark Energy Survey (DES) collaboration. Funding for the DES Projects has been provided by the U.S. Department of Energy, the U.S. National Science Foundation, the Ministry of Science and Education of Spain, the Science and Technology Facilities Council of the United Kingdom, the Higher Education Funding Council for England, the National Center for Supercomputing Applications at the University of Illinois at Urbana-Champaign, the Kavli Institute of Cosmological Physics at the University of Chicago, Center for Cosmology and Astro-Particle Physics at the Ohio State University, the Mitchell Institute for Fundamental Physics and Astronomy at Texas A\&M University, Financiadora de Estudos e Projetos, Fundacao Carlos Chagas Filho de Amparo, Financiadora de Estudos e Projetos, Fundacao Carlos Chagas Filho de Amparo a Pesquisa do Estado do Rio de Janeiro, Conselho Nacional de Desenvolvimento Cientifico e Tecnologico and the Ministerio da Ciencia, Tecnologia e Inovacao, the Deutsche Forschungsgemeinschaft and the Collaborating Institutions in the Dark Energy Survey. The Collaborating Institutions are Argonne National Laboratory, the University of California at Santa Cruz, the University of Cambridge, Centro de Investigaciones Energeticas, Medioambientales y Tecnologicas-Madrid, the University of Chicago, University College London, the DES-Brazil Consortium, the University of Edinburgh, the Eidgenossische Technische Hochschule (ETH) Zurich, Fermi National Accelerator Laboratory, the University of Illinois at Urbana-Champaign, the Institut de Ciencies de l’Espai (IEEC/CSIC), the Institut de Fisica d’Altes Energies, Lawrence Berkeley National Laboratory, the Ludwig Maximilians Universitat Munchen and the associated Excellence Cluster Universe, the University of Michigan, NSF’s NOIRLab, the University of Nottingham, the Ohio State University, the University of Pennsylvania, the University of Portsmouth, SLAC National Accelerator Laboratory, Stanford University, the University of Sussex, and Texas A\&M University.
BASS is a key project of the Telescope Access Program (TAP), which has been funded by the National Astronomical Observatories of China, the Chinese Academy of Sciences (the Strategic Priority Research Program “The Emergence of Cosmological Structures” Grant \# XDB09000000), and the Special Fund for Astronomy from the Ministry of Finance. The BASS is also supported by the External Cooperation Program of Chinese Academy of Sciences (Grant \# 114A11KYSB20160057), and Chinese National Natural Science Foundation (Grant \# 12120101003, \# 11433005).
The Legacy Survey team makes use of data products from the Near-Earth Object Wide-field Infrared Survey Explorer (NEOWISE), which is a project of the Jet Propulsion Laboratory/California Institute of Technology. NEOWISE is funded by the National Aeronautics and Space Administration.
The Legacy Surveys imaging of the DESI footprint is supported by the Director, Office of Science, Office of High Energy Physics of the U.S. Department of Energy under Contract No. DE-AC02-05CH1123, by the National Energy Research Scientific Computing Center, a DOE Office of Science User Facility under the same contract; and by the U.S. National Science Foundation, Division of Astronomical Sciences under Contract No. AST-0950945 to NOAO.

\bibliographystyle{aa}
\bibliography{refs}

\appendix

\section{Technical details of the difference imaging}
\label{sec:method-details}
\subsection{Choice of the reference image}
\label{sec:refimage}
It would be optimal for the transition kernel to be a low-pass filter. 
Indeed in the opposite case, the reference image is convoluted with a high-pass filter 
which equates to a deconvolution. 
This is not necessarily unwanted as the operation is still photometric, 
but it will create ringing artifacts and instabilities. 
Thus when possible, the reference image is chosen to be the one with the best seeing 
such that it needs be blurred to be adapted onto the other images of the set. 
Because of the nature of Pan-STARRS data however, it is not always possible: 
when a large portion of the best seeing image is missing (see section \ref{sec:masking}), 
two or more images (in order of ascending seeing) are combined for the sake of having enough reference stars. 
That is because having too few eligible pixels for the matching procedure decreases 
the condition number of the matrix of Eq.~(\ref{eq:bigmatrix}), 
leading to more numerical instability.

\subsection{Masking bad and irrelevant pixels}
\label{sec:masking}
The matrix in Eq.~(\ref{eq:bigmatrix}) can become very large, 
making the least square problem slower to solve. 
The complexity of this step is roughly $\sim\mathcal{O}\big(N^4(n + N^2)\big)$. 
$N$ is defined by the size of the kernel, but $n$ can be advantageously reduced. 
Indeed most of the pixels in an astronomical image are dark and do not carry information. 
Worse, the brightest stars in a field are usually burnt, 
contribute a lot to the $\chi^2$ due to their high intensity and of course, 
do not contain valid information about the PSF. 
The first step is therefore masking the background and the very bright pixels. 
These were identified using \textit{sextractor} \citep{sextractor}, 
which can produce segmentation maps labelling each star and the background, 
together with the output catalogue of stars linking each label in the segmentation map with a magnitude. 
All stars brighter than a magnitude threshold are masked.
\begin{figure}
\centering
\includegraphics[width=\columnwidth]{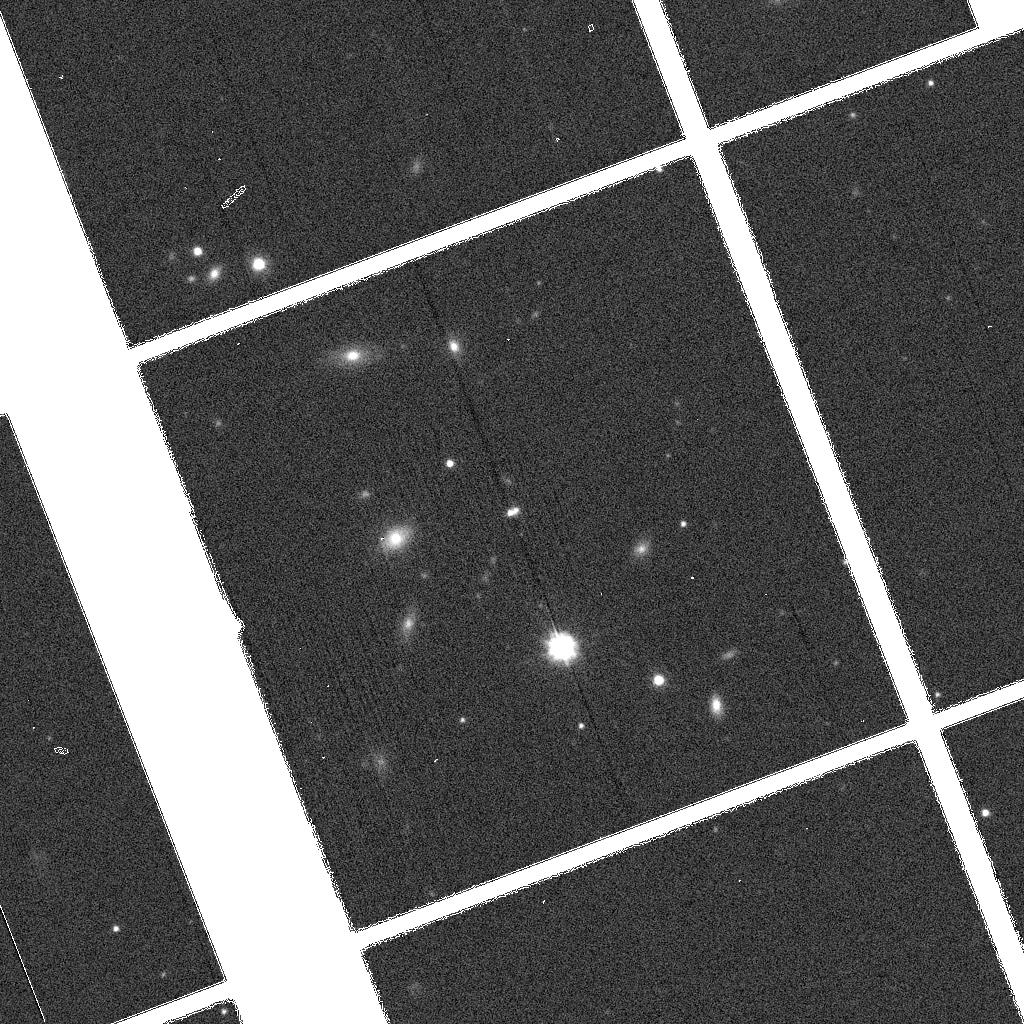}
\caption{
Typical cutout as delivered by the Pan-STARRS cutout server. 
The size of this field is
256\arcsec$\times$256\arcsec (1024$\times$1024 pixels$^2$).
Throughout this work we mostly 128\arcsec$\times$128\arcsec cutouts but allow for the possibility to increase the size of the field when we fail to find useful stars that can constrain the transition kernel.
}
\label{fig:cutout}
\end{figure}
\noindent
Next comes the Pan-STARRS specific masking: 
Pan-STARRS cutouts come with useful auxiliary data such as noise maps 
(used for the weighting of the pixels in the $\chi^2$ minimization) and masks. 
The latter contain flags labelling each pixel with one of 17 ways 
(as defined by Pan-STARRS) a pixel can be bad. 
Fig.~\ref{fig:cutout} shows a typical cutout where empty, 
dead or burnt pixels are visible. 
We only select the pixels that come flag-free. 
Overall, masking decreases $n$ by up to 3 orders of magnitude and substantially increases the robustness of the algorithm.

\subsection{Noise map of a difference image}
\label{sec:noisemap_diffimg} 
When blurring the reference image in particular as in Eq. (\ref{eq:convpixel}) 
the transition kernel $K_j$ was chosen in the sense of the least squares of the sum of Eq. (\ref{eq:chi2})
so as to blur the reference $R(\vec{x})$ to match the resolution of the target image $I(\vec{x})$. 
However the weights $\sigma_R^2(\vec{x})+\sigma_I^2(\vec{x})$ 
do not constitute a good noise map for the resulting difference image 
$\Delta I(\vec{x}) = R_\mathrm{blurred}(\vec{x}) - I(\vec{x})$. 
Indeed the latter is a sum of PSF-matched images, 
whereas the former combines two different resolutions.
As a first order correction, we blur $\sigma_R^2(\vec{x})$ with the 
optimized kernel $K_j$:
\begin{equation}
\sigma^2_{R,\,\mathrm{blurred}}(x_i) = \sum_{j=1}^{N^2} K_j \big(\sigma^2_{R}\big)_{ij}\;.
\end{equation}
where including the background $B$ would not make
sense since the variance is translation invariant.
Then the final noise map is:
\begin{equation}
\label{eq:diffimgnoisemap}
\sigma^2_{\Delta I}(\vec{x}) = \sigma^2_{R,\,\mathrm{blurred}}(\vec{x}) + \sigma^2_I(\vec{x}).
\end{equation}
After the algorithm completes, a new value of reduced $\chi^2$ is calculated using the corrected noise map (\ref{eq:diffimgnoisemap})
and later used for the selection of the best difference images.
With this correction, the reduced $\chi^2$ value averages roughly 1 for 
kernel sizes going from 7$\times$7 to 13$\times$13 pixels$^2$.

\subsection{Choice of the contributing difference images}
\label{sec:choicediff}
When building variability maps, we need to maximize the 
chance that some variability is observed. 
Thus we enforce that exactly one difference image per available night 
of observation is chosen.
Amid the available images in each night, we choose the one which has the 
best covering of the central region. 
Should all of them offer a decent covering, we select the one with the best seeing value.
Naturally the difference images whose generation procedure failed 
(reduced $\chi^2$ over some threshold) are not candidates to 
this heuristic of selection.

\subsection{Kernel extent and overfitting}
Matching two images with a large difference in their PSF requires 
a transfer kernel with a large spatial extent. 
In this implementation, extending the kernel by a factor $Q$ makes the number 
of degrees of freedom in the fits $Q^2$ folds larger. 
This often led to reduced $\chi^2$ values much smaller than 1, 
suggesting that the fit could be carried out meaningfully 
and faster with fewer degrees of freedom. 
In the present context things were unfolding fast enough 
and no further optimization was undertaken. 
When working with very large images however, 
one might want to implement the multi-resolution approach 
developed by \citet{Bramich2012}. 
In this paradigm, the outer pixels of the kernel are binned as to increase 
the spatial extent of the kernel without quadratically 
increasing the number of degrees of freedom.

\subsection{Examples of corrupted variability maps}
\label{sec:broken}
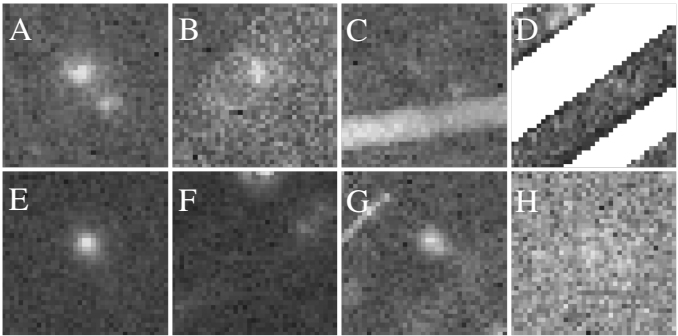
\begin{figure}[ht!]
\resizebox{\columnwidth}{!}{%
\begingroup%
  \makeatletter%
  \providecommand\color[2][]{%
    \errmessage{(Inkscape) Color is used for the text in Inkscape, but the package 'color.sty' is not loaded}%
    \renewcommand\color[2][]{}%
  }%
  \providecommand\transparent[1]{%
    \errmessage{(Inkscape) Transparency is used (non-zero) for the text in Inkscape, but the package 'transparent.sty' is not loaded}%
    \renewcommand\transparent[1]{}%
  }%
  \providecommand\rotatebox[2]{#2}%
  \newcommand*\fsize{\dimexpr\f@size pt\relax}%
  \newcommand*\lineheight[1]{\fontsize{\fsize}{#1\fsize}\selectfont}%
  \ifx\svgwidth\undefined%
    \setlength{\unitlength}{205.87591697bp}%
    \ifx\svgscale\undefined%
      \relax%
    \else%
      \setlength{\unitlength}{\unitlength * \real{\svgscale}}%
    \fi%
  \else%
    \setlength{\unitlength}{\svgwidth}%
  \fi%
  \global\let\svgwidth\undefined%
  \global\let\svgscale\undefined%
  \makeatother%
  \begin{picture}(1,0.49866522)%
  \centering%
    \lineheight{1}%
    \setlength\tabcolsep{0pt}%
    \put(0,0){\includegraphics[width=\unitlength,page=1]{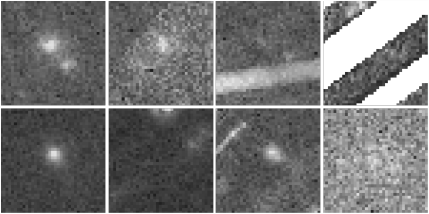}}%
    \put(0.01021814,0.4420404){\color[rgb]{1,1,1}\makebox(0,0)[lt]{\lineheight{1.25}\smash{\begin{tabular}[t]{l}A\end{tabular}}}}%
    \put(0.2614214,0.4420404){\color[rgb]{1,1,1}\makebox(0,0)[lt]{\lineheight{1.25}\smash{\begin{tabular}[t]{l}B\end{tabular}}}}%
    \put(0.5093816,0.44138984){\color[rgb]{1,1,1}\makebox(0,0)[lt]{\lineheight{1.25}\smash{\begin{tabular}[t]{l}C\end{tabular}}}}%
    \put(0.75978603,0.4420404){\color[rgb]{1,1,1}\makebox(0,0)[lt]{\lineheight{1.25}\smash{\begin{tabular}[t]{l}D\end{tabular}}}}%
    \put(0.01021814,0.1883335){\color[rgb]{1,1,1}\makebox(0,0)[lt]{\lineheight{1.25}\smash{\begin{tabular}[t]{l}E\end{tabular}}}}%
    \put(0.2614214,0.18775585){\color[rgb]{1,1,1}\makebox(0,0)[lt]{\lineheight{1.25}\smash{\begin{tabular}[t]{l}F\end{tabular}}}}%
    \put(0.5093816,0.18660863){\color[rgb]{1,1,1}\makebox(0,0)[lt]{\lineheight{1.25}\smash{\begin{tabular}[t]{l}G\end{tabular}}}}%
    \put(0.75978603,0.18573139){\color[rgb]{1,1,1}\makebox(0,0)[lt]{\lineheight{1.25}\smash{\begin{tabular}[t]{l}H\end{tabular}}}}%
  \end{picture}%
\endgroup%

}
\caption{Different crops of variability maps and problems. 
A:~a valid detection. 
B:~one of the original images had normalisation problems. 
C:~CCD bleeding in one of the original images because of a very bright star. 
D:~Heavily banded region, missing data. 
E:~valid image, but only one variable patch. 
F:~The original images contain a very bright star which cannot 
be properly subtracted at the edge. 
G:~An uncaught cosmic.
H:~valid image, no variability.}\label{fig:varmaps}
\end{figure}

The proper classification of the variability maps 
posed a challenge because of the numerous ways the maps can be 
corrupted. 
A few examples, both bad and good are 
listed in Fig.~\ref{fig:varmaps}. 
Sometimes the difference imaging
will simple fail despite the reduced $\chi^2$ values being
in an acceptable range. 
This can arise, for example, when the candidate
is much brighter than any of the stars in the field,
the structure of the PSF is then under-determined and 
the transition kernel will only guarantee the proper
subtraction of sources near the noise level. 
This will also happen with meteors or satellites, 
or in the rarer case where one of the cutouts is not properly 
aligned with the others.

\end{document}